\documentclass[10pt]{article}  
\usepackage{amsmath,amssymb,amsfonts}
\usepackage{graphicx}
\begin{document}

\title{Linear programming bounds for unitary space time codes}

\author{
{Jean Creignou}
\and
 {Herv\'e Diet}
}

\newtheorem{defi}{Definition}[section]
\newtheorem{definition}[defi]{Definition}
\newtheorem{proposition}[defi]{Proposition}
\newtheorem{theorem}[defi]{Theorem}
\newtheorem{remark}[defi]{Remark}
\newtheorem{corollary}[defi]{Corollary}
\newtheorem{lemma}[defi]{Lemma}

\newcommand {\F}{\mathbb{F}}
\newcommand {\G}{\mathcal{G}}
\newcommand {\ie} {\textit{i.e.}}
\newcommand {\trace}{\textrm{Trace}}
\newcommand {\conj}[1] {\overline{#1}}
\newcommand {\Chi} {\chi}
\newcommand{\Nu}{\mathcal V}
\newcommand{\N}{\mathbb{N}}
\newcommand{\Z}{\mathbb{Z}}
\newcommand{\R}{\mathbb{R}}
\newcommand{\C}{\mathbb{C}}
\newcommand{\U}{\mathbb{U}}
\newcommand{\PS}{\mathcal{P}}
\newcommand{\QS}{\mathcal{Q}}
\newcommand{\Code}{\mathcal V} 
\newcommand{\Cl}[2]{ \textrm{Cl}_{#1}({#2})}

\newenvironment{proof}{\noindent {\bf Proof :}}{}

\maketitle

\begin{abstract} 
The linear programming method is applied to the space $\U_n(\C)$ of
unitary matrices  in order 
 to obtain bounds for codes relative to the diversity sum and the diversity product.
 Theoretical and numerical results improving previously known bounds are derived.
 \end{abstract}

\section{Introduction}
Nowadays breakthrough of wireless communications has provided new and nice 
 problems to the field of coding theory. Indeed strategic issues of MIMO communications has
 lead to consider coding in Grassmann and Stiefel manifolds, and also
 in unitary groups \cite{TSE}.
 Codes in unitary groups  are useful in the context of non-coherent flat Rayleigh channel as shown
 in \cite{MARZETTA}. The performance of a unitary space-time code $\Code$
 is measured  (see \cite{ROSEN}) by two functions namely the diversity sum $\Sigma \Nu$ and the 
 diversity product $\Pi \Nu$ : 
 \begin{eqnarray*}
 \Sigma \Nu := \frac{1}{2\sqrt{n}} \min \left \{ ||x-y||\ :
 \ x\ne y \in \Code   \right\},
 \end{eqnarray*}
 \begin{eqnarray*}
 \Pi \Nu := \frac{1}{2} \min \left \{ |\det(x-y)|^{\frac 1 n}\ :
 \ x\ne y \in \Code \right\}.
 \end{eqnarray*}
Here $||A||$ denotes the standard Euclidean norm of complex matrices: $||A||^2=\trace(AA^*)=\sum|A_{i,j}|^2$.
A standard problem is, given a  number $N$ of points in
$\U_n(\C)$, to maximize the value of $\Sigma \Nu$ or of $\Pi \Nu$.
Many authors have addressed this question,
 narrowing gaps between bounds and explicit constructions (see
 \cite{ROSEN}, \cite{AB}, \cite{RECITE},  \cite{SHOKRO}). The
 linear programming method, which was initially developed by Philippe
 Delsarte in the framework of association schemes \cite{DELSARTE}, and
 is
 a powerful method to deal with such questions, has not been applied
 to unitary codes before.
 Delsarte method was successfully adapted to the compact two-point homogeneous
 spaces by Kabatiansky and Levenshtein \cite{LEVEN} and  recently to
 more general situations like the Grassmann codes \cite{CHRISTINE1},
 the permutation codes \cite{TARNANEN}, the ordered codes \cite{BARG},
 \cite{BIERBRAUER}. Most of the
 situations mentioned above fit into a common framework, namely a
 compact group $G$ acts homogeneously on the underlying space $X$, and the
 representation theory of $G$ constructs a certain family of orthogonal
 polynomials naturally attached to $X$. Standard methods of harmonic
 analysis show that these polynomials hold the desired positivity
 property that allows for Delsarte linear programming method. This
 general framework is recalled in Section II, and we show in  Section III
 that unitary codes can be treated likewise, the  Schur
 polynomials being the associated family of orthogonal
 polynomials. Section IV and V present the results, both numerical and
 analytic, obtained by the implementation of this method. 
 It turns out that we  improve all previously known bounds concerning the diversity sum and 
 the diversity product. Moreover it is worth pointing out that the mathod can easily be extended
 to more complex situations, for example a diversity function involving 
 both  $\Sigma \Nu$ and $\Pi \Nu$.\\
 
 \section{The linear programming method on homogeneous spaces} 
 
We briefly describe the linear programming method on homogeneous
spaces. For more details we refer to \cite{LEVEN}, \cite[Chapter 9]{CS} for a
treatment of 2-point homogeneous spaces, and to \cite{CHRISTINE1} for
the prominent case of Grassmann codes. 
  
 Let $G$ be a compact group acting transitively and continuously on a
 compact space $X$, and $\tau : X\times X  \to Y$ such that $\tau$ 
 characterizes the orbits of $G$ acting on $X\times X$. We mean here that, for all $x_1,x_2,x'_1,x'_2 \in X$,
\begin{equation*}
\tau(x_1,x_2)=\tau(x'_1,x'_2) \Leftrightarrow \exists  g \in G :
g(x_1,x_2)=(x'_1,x'_2).
\end{equation*}
     
  Let $S$ be any subset of $Y$, we call a finite subset $\Code \subset
  X$ a $S$-code if for all
  $c_1 \ne c_2 \in \Code,\ \tau(c_1,c_2) \in S$. 
   
 A continuous function $P : Y \to \C$ is said to possess the positivity property 
 if for any finite subset
 $\Code \subset X$ and any complex function $\alpha : X \to \C$, 
 
  $$\sum_{x,y \in \Code} \alpha(x) \conj{\alpha(y)} P\left(\tau(x,y)\right)
  \textrm{ is real non-negative.}$$
  A canonical example is  the constant function $P_0=1$; non-trivial 
  examples are given by the so called zonal functions that we
  introduce now. 

Let   $L^2(X)=\left\{ u:X\to\C : \int_X |u(x)|^2 dx < \infty\right\}$ where
  $dx$ is the unique $G$-invariant Haar measure on $X$ such that $\int_X  dx =1$.
  This vector space is given the standard $G$-action defined by
  $g.u(x)=u(g^{-1}(x))$ and is endowed with the canonical $G$-invariant hermitian 
  product : $(u_1,u_2)=\int_X u_1(x)\conj{u_2(x)} dx.$ 
The Peter-Weil theorem shows that $L^2(X)$ can be decomposed as a
direct sum of 
  $G$-irreducible subspaces $V_i$. The next step is to associate to
  each  irreducible subspace $V_i$ a so-called zonal function
  $P_{V_i}$. A standard construction is the following: given an
  orthonormal  basis $(u_1(x),...,u_d(x))$ of $V_i$,  one can define 
   \begin{eqnarray}\label{DEFEQ}
   \tilde P_{V_i}(x,y)= \frac 1 d_i \sum_{i=1}^{d_i} u_i(x)\conj{u_i(y)}.
   \end{eqnarray}
where $d_i=\dim(V_i)$. 
Since  these functions are constant on $G$-orbits we can rewrite
   $\tilde P_{V_i}(x,y)=P_{V_i}\left(\tau(x,y)\right)$. From this property 
   comes the term \emph{zonal functions} used to qualify them. 
   From equation (\ref{DEFEQ}) it is easy to prove that these zonal
   functions verify the positivity property and do not depend of the chosen orthonormal basis. 
It turns out that, when the irreducible subspaces $V_i$ are pairwise
non isomorphic,  the cone of continuous positive $G$-invariant functions is exactly
the set of linear combinations with non negative coefficients of the
$P_{V_i}$
(see \cite{BOCHNER}). In the remaining of this paper we assume that
this condition is satisfied. We moreover let $V_0$ denote the
one-dimensional subspace associated to the trivial representation of $G$.

  The so-called linear programming bounds are obtained with  the following theorem :

 \begin{theorem}\label{FUNDATHM} Let $P=\sum_{i} c_i P_{V_i}$ 
 a linear combination of the zonal functions $P_{V_i}$ with a finite
 number of non zero coefficients.
  Assume furthermore  that : $c_i\ge 0$, $c_0 >0$ and $\Re(P)$ (the
  real part of $P$) is
 non-positive on $S$. Then any $S$-code verifies
  \begin{eqnarray}\label{FUNDA} |\Code| \le \frac{ P(\tau_0)}{c_0}
   \end{eqnarray}     
   where $\tau_0=\tau(x,x)$ for any $x$. 
\end{theorem}
   
   \begin{proof} On one hand,
\begin{align*}
   \sum_{x,y \in \Code} P(\tau(x,y))&= \sum_{x=y \in \Code}
   P(\tau(x,y)) + \sum_{x\ne y \in \Code} P(\tau(x,y))\\
 &\leq |\Code|P(\tau_0) 
\end{align*}
On the other hand,
\begin{align*}
\sum_{x,y \in \Code} P(\tau(x,y)) &= \sum_{x,y \in \Code} c_0 P_0(\tau(x,y)) +
\textrm{non negative terms} \\
&\ge |\Code|^2c_0.
\end{align*}
   \end{proof}

 \section{The  case of  unitary codes }
  As a particular case we set $G=\U_n(\C)\times \U_n(\C)$ and $X=\U_n(\C)$. For $g=U\times V$ and
  $x=M$ we set $gx=UMV^{-1}$. In this context the orbit of a pair $(x,y)$ is characterized by
  the eigenvalues $(e^{i\theta_1},...,e^{i\theta_n})$ of the unitary matrix $xy^{-1}$. 
  The Peter-Weil theorem gives a decomposition of $L^2(X)$ 
  into irreducible subspaces 
\begin{equation}\label{dec}
L^2(X)=\oplus (V_{\chi}\otimes V_{\conj{\chi}})
\end{equation}
where the sum  runs over all irreducible representations $V_{\chi}$ of
$\U_n(\C)$. It is worth noticing that the $G$-subspaces $V_{\chi}\otimes V_{\conj{\chi}}$ are pairwise non isomorphic.  From this decomposition
  one can deduce the following theorem :

  \begin{proposition}
  The zonal functions associated to this decomposition are
\begin{equation*}
 P_\chi(x,y)=\chi(xy^{-1})
\end{equation*}
where $\chi$ denotes any \textrm{ irreducible character of } $\U_n(\C)$.
  \end{proposition}

  The irreducible characters of $\U_n(\C)$ are known to be finite
  dimensional and to have a nice description using Schur polynomials (\cite{FULTON}). 
 We recall briefly some notations and definitions concerning those polynomials.

 For any integer $k$, a partition of $k$ in $n$ parts is a finite decreasing sequence
 of $n$ non negative integers which sum exactly to $k$ ($k$ is also called the degree
 of the partition). A partition $\lambda$ is dominating $\mu$ (noted $\lambda \succ \mu$)
 if $\forall r\le n,\ \sum_{i=1}^r \lambda_i \ge \sum_{i=1}^r \mu_i$. Given a partition 
 $\lambda=[\lambda_1,...,\lambda_n]$ we define the elementary symmetric polynomials
 $m_\lambda \in \Z[x_1,...,x_n]$ as  the renormalization to a monic polynomial
 of $$\sum_{\sigma \in \mathfrak{S}_n}  x_{\sigma(1)}^{\lambda_1}x_{\sigma(2)}^{\lambda_2}...x_{\sigma(n)}^{\lambda_n}.$$
 These polynomials form a basis of the set of symmetric polynomials $\Z[x_1,...,x_n]^{\textrm {Sym}}$.
 Schur polynomials have been intensively studied and have several definitions. 
 For our purpose we will define the Schur polynomials $S_\lambda$ as 
 $$S_\lambda = \sum_{\lambda \succ \mu} K_{\lambda,\mu} m_\mu,$$
 where the $K_{\lambda,\mu} \in \N$ are the so called Kostka numbers.
 For more precision on those numbers see \cite{SAGAN}.
 It is clear that Schur polynomials form another basis 
 of $\Z[x_1,...,x_n]^{\textrm {Sym}}$.

It is well known that the irreducible polynomial characters of
$\U_n(\C)$ are expressed using Schur polynomials 
 (we refer to \cite{FULTON} for details) in the following way:
 let $(e^{i\theta_1},...,e^{i\theta_n})$ denote the eigenvalues of a
 unitary matrix $M$, and $\lambda$ a partition. Then
\begin{equation*}
\chi_\lambda(M)=S_\lambda(e^{i\theta_1},...,e^{i\theta_n}). 
\end{equation*}
One obtains all irreducible  characters of $\U_n(\C)$ by multiplying
the characters $\chi_{\lambda}$ by a relative power of $\det(M)=\prod
e^{i\theta_k}$.
All together, we obtain the theorem:

\begin{theorem}
For all partition $\lambda$ and $s\in \Z$, let
\begin{equation}\label{P1}
P_{\lambda,s}(x_1,\dots,x_n)=(x_1\dots x_n)^sP_{\lambda}(x_1,\dots, x_n).
\end{equation}
These rational fractions give the zonal functions associated to the
irreducible decomposition \eqref{dec} in the following way:
if $\chi\simeq \det^s\otimes\lambda$, if the eigenvalues of $xy^{-1}$ are
$(e^{i\theta_1},...,e^{i\theta_n})$, then 
\begin{equation*}
P_{\chi}(x,y)=P_{\lambda,s}(e^{i\theta_1},...,e^{i\theta_n}).
\end{equation*}
\end{theorem}.

 We are now almost ready to compute bounds for unitary codes $\Code \subset \U_n(\C)$.
 Let
 $$d_\Sigma^2(e^{i\theta_1},...,e^{i\theta_n}):=\frac{1}{2n} {\sum_{i=1}^n (1-\cos \theta_i)}$$
 and 
 $$d_\Pi^2(e^{i\theta_1},...,e^{i\theta_n}):=\frac{1}{2}  \left(\prod_{i=1}^n (1-\cos \theta_i) \right)^{\frac{1}{n}}.$$
 and define $d_\Sigma^2(x,y)$ (resp. $d_\Pi^2(x,y)$) to be the above functions evaluated at the 
 eigenvalues of $xy^{-1}$. 
 These functions are related to the diversity sum and diversity product by 
 $$\Sigma \Nu = \min_{\substack{x,y \in \Code\\x\ne y}} d_\Sigma(x,y) \textrm{ and }
 \Pi \Nu = \min_{\substack{x,y \in \Code\\x\ne y}} d_\Pi(x,y).$$
 We recall that the orbit of a pair $(x,y)$ is characterized by the eigenvalues 
 of $xy^{-1}$, so $d_\Sigma$ and $d_\Pi$ are constant on $G$-orbits. 
 We may now define, with the notations of Section II,  the sets $S$ related to each diversity function:
 $$S_\Sigma(\delta):=\{ (e^{i\theta_1},...,e^{i\theta_n}) :  d_\Sigma(e^{i\theta_1},...,e^{i\theta_n}) \ge \delta \}$$ 
 $$S_\Pi(\delta):=\{ (e^{i\theta_1},...,e^{i\theta_n}) :  d_\Pi(e^{i\theta_1},...,e^{i\theta_n}) \ge \delta \}.$$

\section{Analytic bounds of low degree}

 From the explicit description of the zonal functions \eqref{P1},
 we have deduced convenient 
polynomials of low degree which verify the positivity property. Using 
formula (\ref{FUNDA})  we derive the analytic bounds of Theorems \ref{THM1} and \ref{THM2}.\\

  \noindent\begin{theorem}\label{THM1}
  Let $\Code$ be a unitary space time code with diversity sum $\Sigma \Nu$, the following upper bounds hold : 
  \noindent \begin{equation}
     \noindent  \quad |\Code|\le  \frac{2(\Sigma \Nu)^2}{2(\Sigma \Nu)^2-1},
    \quad \text{if }(\Sigma \Nu)^2 >\frac{1}{2}
    \end{equation}
    \begin{equation}\label{DEG2SUM}
   \quad |\Code| \le \frac{8n^2(\Sigma \Nu)^2}{4n^2(\Sigma \Nu)^2-(2n^2-1)}, 
     \text{if }(\Sigma \Nu)^2 > \frac{2n^2-1}{4n^2}
    \end{equation}
    \begin{equation}
   \quad |\Code| \le \frac{16 n^2 (\Sigma \Nu)^2} {2n(2n-1)(\Sigma \Nu)^2-(2n^2-n-2)},\end{equation}
   $$\\ \quad \text{if } (\Sigma \Nu)^2 \ge \frac{2n^2-n-2}{2n(2n-1)}$$\\
 \end{theorem}

\noindent \begin{theorem}\label{THM2}
Let $\Code$ be a unitary space time code with diversity product $\Pi \Nu$, the following upper bounds hold :
 \begin{equation}
   |\Code| \le \frac{2(\Pi \Nu)^2}{2(\Pi \Nu)^2-1}, \quad \text{if }(\Pi \Nu)^2 > \frac{1}{{2}}
 \end{equation}   
 
\begin{equation}\label{DEG2PROD}
   |\Code| \le \frac{8n(\Pi \Nu)^2 }{4n(\Pi \Nu)^2-(2n-1) },  \text{if }(\Pi \Nu)^2 > \frac{2n-1}{4n}, \ n \ge 3
 \end{equation}   
 
\begin{equation}
   |\Code| \le \frac{8(\Pi \Nu)^6+4(\Pi \Nu)^4+8(\Pi \Nu)^2}{8(\Pi \Nu)^6-\frac{1}{4}},
 \end{equation}\\
$$\text{if }(\Pi \Nu)^2 \ge \frac{1}{2}, \ n=2.$$\\
 \end{theorem}

 The proofs of these theorems are based on Theorem \ref{FUNDATHM} and
 on  the following lemma : 

 \begin{lemma} Let $y_j=\cos \theta_j$ and 
  $m_{[a_1,...,a_r]}(y)$ the elementary symmetric
  polynomials in  the $y_j=\cos \theta_j$.
  The following polynomials  are linear combination of the zonal
  functions \eqref{P1} with non negative coefficients:
  $$\begin{array}{l}  Q_{[0]}=1 \\
 Q_{[11]}=m_{[11]}(y)+\frac{(n-1)}{4}\\
 Q_{[1]}=m_{[1]}(y) \\
 Q_{[2]}=m_{[2]}(y)+m_{[11]}(y)-\frac{(n+1)}{4}\\
  Q_{[1,1,1]} =\ m_{[1,1,1]}(y) + \frac{(n-2)}{4}m_{[1]}(y)\\
 Q_{[2,1]} =\  m_{[2,1]}(y) +2 m_{[1,1,1]}(y) -\frac{1}{4} m_{[1]}(y) \\
 Q_{[3]} = \  m_{[3]}(y)+m_{[2,1]}(y)+m_{[1,1,1]}(y) -\frac{(n+2)}{4} m_{[1]}(y)\\
  \end{array}$$\\
  \end{lemma}

 \begin{proof}[Theorem \ref{THM1}, Sketch] 
    Let $s=\frac 1 n \sum_{j=1}^n \cos \theta_j$ so that  $s=1-2d_\Sigma^2$.
   We apply formula (\ref{FUNDA}) with the polynomials :
   $P=Q_{[1]}-ns$, 
   $P=(Q_{[1]}-ns)(Q_{[1]}+n)$, $P=(Q_{[1]}-ns)R$, where $R$ is the symmetrization of
   $(y_1+1)(y_2+1)$.

 \end{proof}

 \begin{proof}[Theorem \ref{THM2}, Sketch] 
Let $p=2d_\pi^2$. We apply formula (\ref{FUNDA}) with the polynomials : 
 $P=Q_{[1]}-n(p-1)$, $P=Q_{[2]}+\frac{(n+1)p}{2}Q_{[1]}+\frac{(n+1)(2n(p-1)+1)}{4}$ \quad (if $n\ge 3$),\\  
 $P=Q_{[2]}+(\frac{p^2}{2}+2p-1)Q_{[1]}+(p^3-\frac{1}{4})$ \quad (if $n=2$).\\
 \end{proof}

 \section{Numerical bounds}
  Numerical
  programs give accurate approximations of the best linear programming
  bounds over a large interval of validity, not covered by the bounds
  proved in Section IV. 
The following curves plot  the linear programming bound on the
diversity functions as a function of the cardinality of the code.
The programs optimize the choice of a polynomial in the variables
$(\cos\theta_1,\dots,\cos\theta_n)$, with degree at most equal to some
parameter $D$.  Increasing $D$ gives
accurate results over a wider range of values for the  diversity
functions, but also increase the computational time.
We use $D=19$ for $n=2$ and $D=13$ for $n=3$.
 \noindent\includegraphics[width=6.2cm,angle=-90]{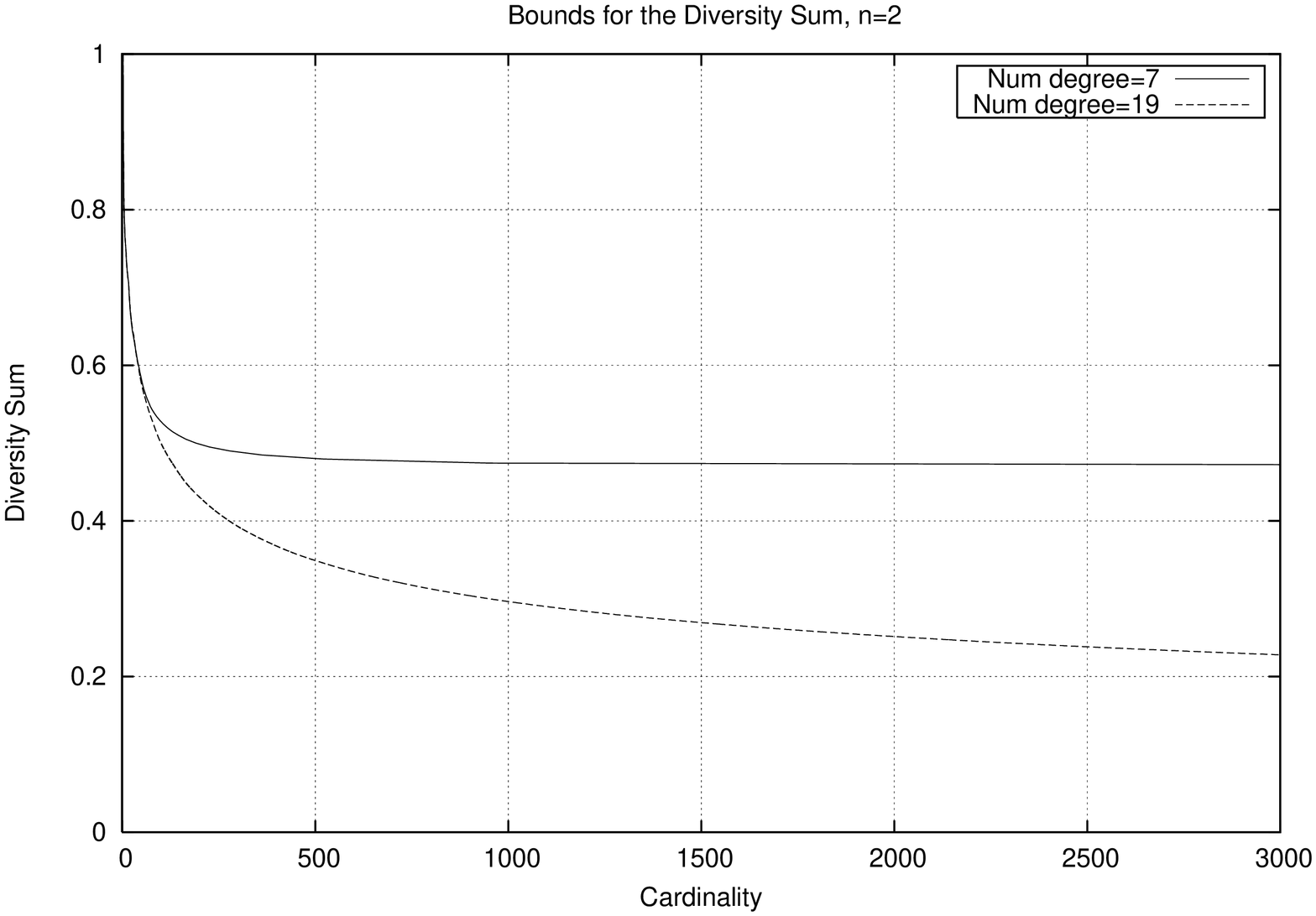}\\ 
\includegraphics[width=6.2cm,angle=-90]{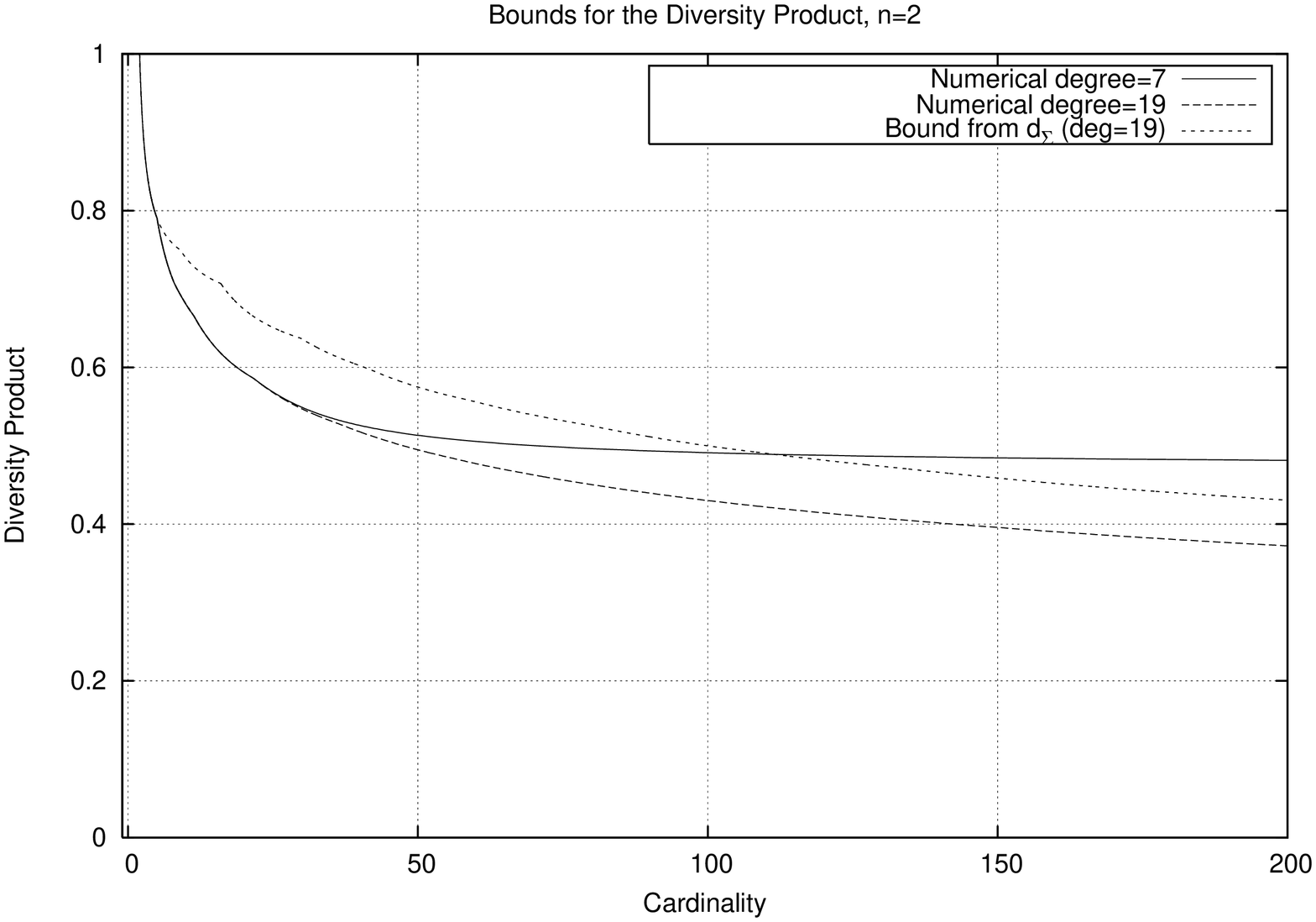}\\ 
\includegraphics[width=6.2cm,angle=-90]{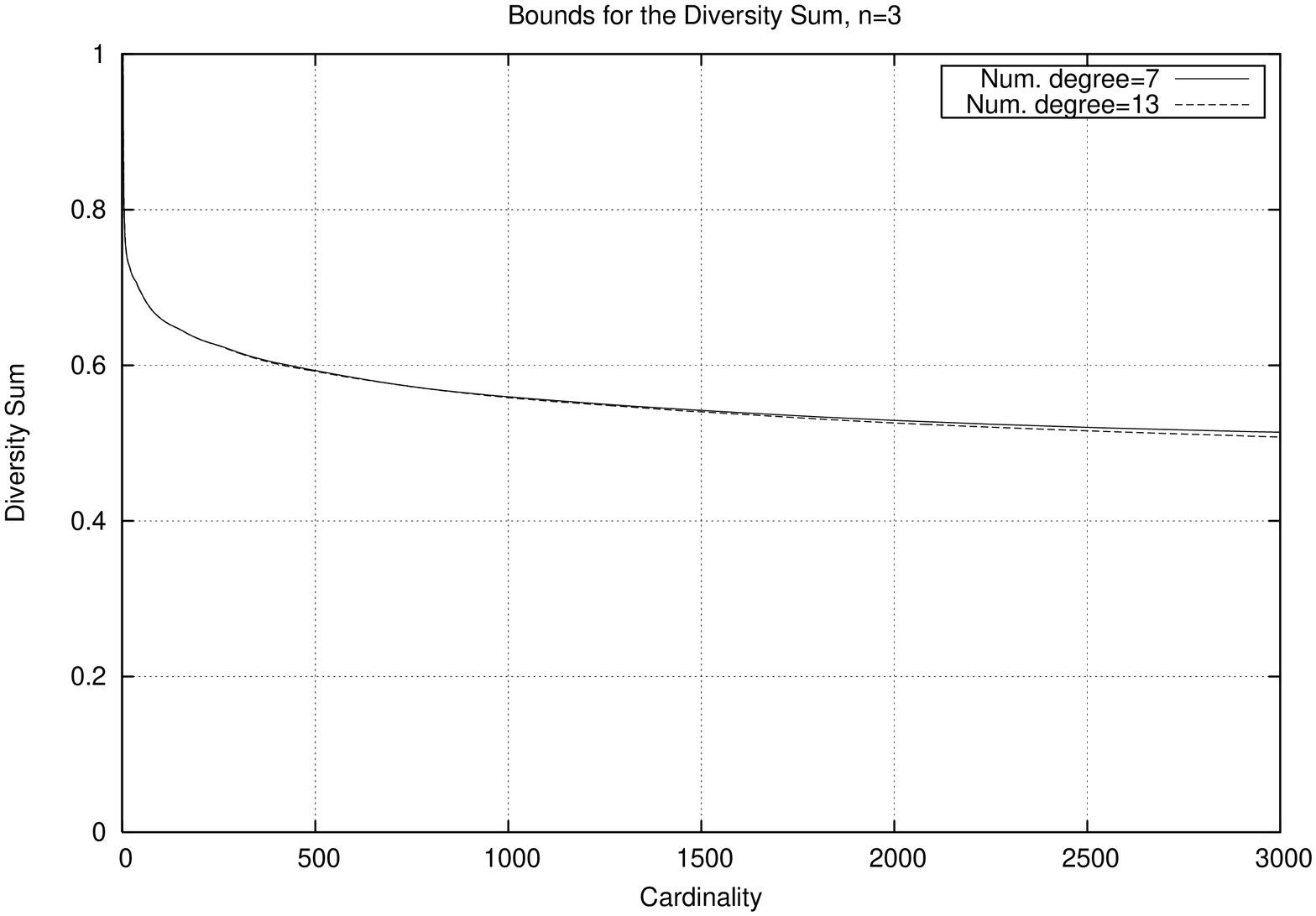}\\
 \includegraphics[width=6.2cm,angle=-90]{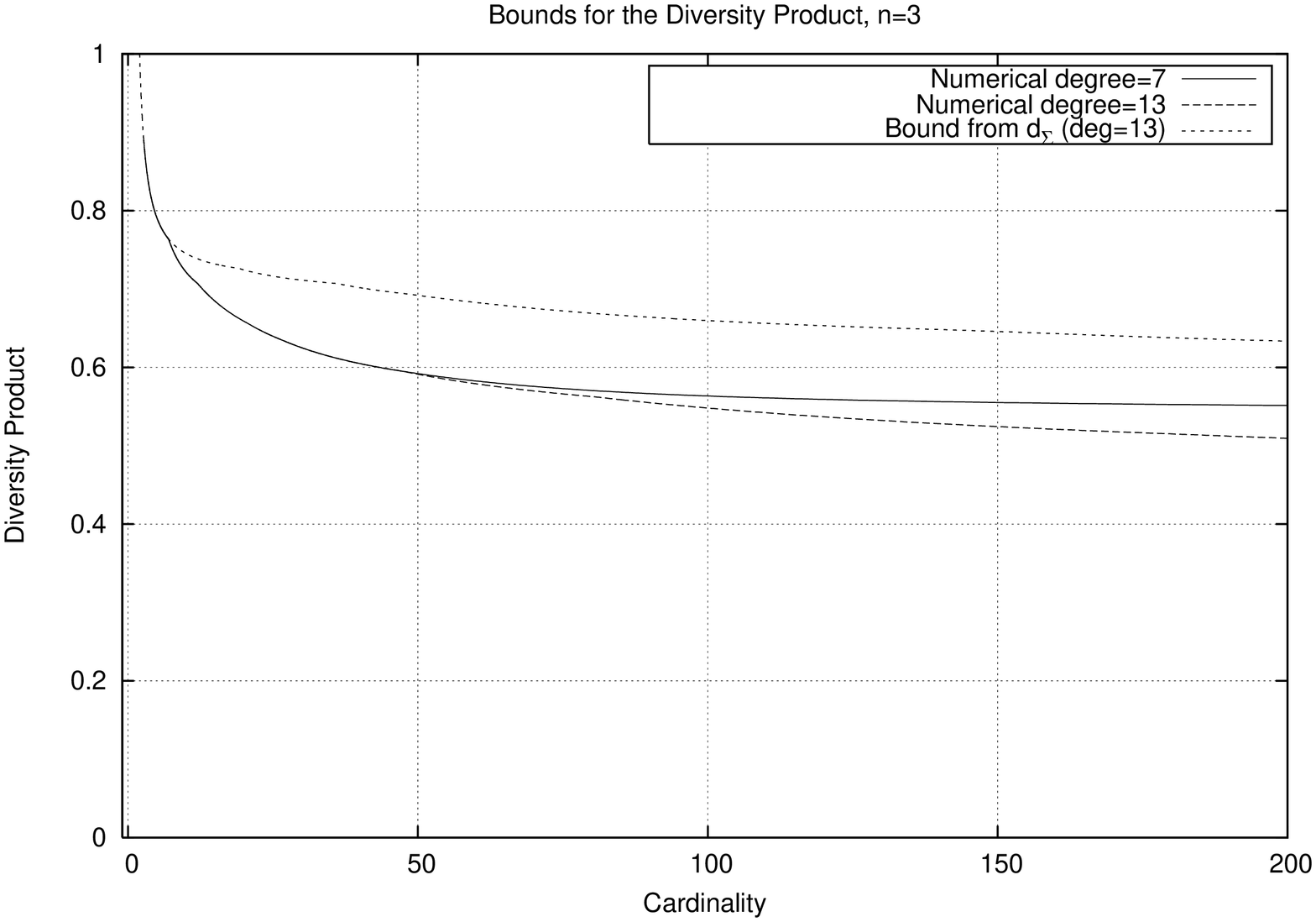}\\

The Tables I and II compare the linear programming bounds for the
dimensions $2$ and $3$ with the 
previous results of \cite{RECITE} and \cite{ROSEN} and show an
improvement in all cases.

 These tables give upper bounds for the  diversity when the
 cardinality $N$ is fixed. 
 All entries except the ones of the last line (LP $d_\Pi$) 
 are concerned with  the diversity sum. The second line tabulates the
 bounds settled in  \cite{RECITE}, obtained using
 Coxeter upper bounds. The two bounds of \cite{ROSEN} were obtained using sphere volume computations.

\begin{table}
\noindent \begin{tabular}{|c|c|c|c|c|}
  \hline
  N & 24 & 48 & 64 & 80 \\
  \hline
  bounds in \cite{RECITE} & 0.6746 & 0.6193 & 0.5969 & 0.5799\\
  \hline
  B1 \cite{ROSEN} & 0.7598 & 0.6603 & 0.6131 & 0.5932\\
  \hline
  B2 \cite{ROSEN} & 0.7794 & 0.6734 & 0.6235 & 0.6026\\
  \hline
  LP $d_\Sigma$  & 0.6547 & 0.5797 & 0.5488 & 0.5254 \\
  \hline
  LP $d_\Pi$ & 0.5730 & 0.4989 & 0.4711 & 0.4504\\
  \hline
  \end{tabular}

 \vspace{10pt}

\noindent \begin{tabular}{|c|c|c|c|c|}
  \hline
  N & 100 & 120 & 128 & 1000\\
  \hline
  bounds in \cite{RECITE} & 0.5632 & 0.5499 & 0.5452 &\\
  \hline
  B1 \cite{ROSEN} & 0.5578 & 0.5425 & 0.5347 & 0.3270\\
  \hline
  B2 \cite{ROSEN} & 0.5654 & 0.5496 & 0.5415 & 0.3285\\
  \hline
  LP $d_\Sigma$ & 0.4999 & 0.4816 & 0.4753 & 0.2964\\
  \hline
  LP $d_\Pi$ & 0.4301 & 0.4144 & 0.4089 & 0.2574 \\
  \hline
 \end{tabular}
\caption{$n=2$}
\end{table}

\begin{table}

\noindent \begin{tabular}{|c|c|c|c|c|}
  \hline
  N & 24 & 48 & 64 & 80\\
  \hline
  LP $d_\Sigma$ & 0.7178 & 0.6939  &  0.6797& 0.6692 \\
  \hline
  LP $d_\Pi$ & 0.6431 & 0.5942 &0.5752 &  0.5628   \\
  \hline
 \end{tabular}\\

\vspace{10pt}

\noindent \begin{tabular}{|c|c|c|c|c|}
  \hline
  N & 100 & 120 & 128 & 1000\\
  \hline
  LP $d_\Sigma$ & 0.6598  & 0.6532 & 0.6511 & 0.5586 \\
  \hline
  LP $d_\Pi$ & 0.5482 & 0.5369 &  0.5332 & 0.4330 \\
  \hline
 \end{tabular}
\caption{$n=3$}
\end{table}

 Moreover, 
concerning the  diversity product, both our 
numerical results and analytic results (compare (\ref{DEG2SUM}) and (\ref{DEG2PROD})) show a large
 gap between the bounds for diversity sum and diversity product, in
 favor of the diversity product. This
 is worth to point out since in previous publications
 bounds for the diversity product were essentially  deduced from the
 trivial inequality $\Sigma \Nu \ge \Pi \Nu$ and hence appear to be
 weak.

\section{Conclusions}

 In this paper we have developed  the linear programming method for
 the unitary space time codes. We have obtained both numerical and analytic bounds.
 The results improve previously known bounds. Furthermore the
 linear programming method allows to deal with
 non-distance functions as the diversity product directly.
 
\section*{Acknowledgment}
The authors would like to thank Christine Bachoc for her precious advises on the  writing of this article.

\end{document}